\documentclass[fleqn,usenatbib]{mnras}

\usepackage{txfonts}
\usepackage[T1]{fontenc}
\usepackage{ae,aecompl}

\usepackage{graphicx}
\usepackage{bm}		% Bold maths symbols, including upright Greek
\usepackage[flushleft]{threeparttable}
\usepackage{float}
\usepackage[caption = false]{subfig}

\usepackage{multicol}        % Multi-column entries in tables

\usepackage{pdflscape}
\title[Coronal properties of QSO\,B2202--209]{Coronal properties of the luminous radio-quiet quasar QSO\,B2202--209}

\author[E. S. Kammoun et al.]{E. S. Kammoun,$^{1}$\thanks{E-mail: \href{mailto:ekammoun@sissa.it}{ekammoun@sissa.it}}
G. Risaliti,$^{2,3}$ D. Stern,$^{4}$ H. D. Jun,$^{4}$ M. Graham,$^{5}$ A. Celotti,$^{1,6,7}$
\newauthor
E. Behar,$^{8}$ M. Elvis,$^{9}$ F. A. Harrison,$^{5}$ G. Matt$^{10}$ and D. J. Walton$^{4,5}$
\\
$^{1}$SISSA, via Bonomea 265, I-34135 Trieste, Italy\\
$^{2}$Dipartimento di Fisica e Astronomia, Universit\`{a} di Firenze, via G. Sansone 1, 50019 Sesto Fiorentino (Firenze), Italy\\
$^{3}$INAF - Osservatorio Astrofisico di Arcetri, Largo E. Fermi 5, I-50125 Firenze, Italy\\
$^{4}$Jet Propulsion Laboratory, California Institute of Technology, 4800 Oak Grove Drive, Pasadena, CA 91109, USA\\
$^{5}$California Institute of Technology, 1200 East California Boulevard, Pasadena, CA 91125, USA\\
$^{6}$INAF  -Osservatorio Astronomico di Brera, via Bianchi 46, I-23807 Merate, Italy\\
$^{7}$INFN - Sezione di Trieste, via Valerio 2, I-34127 Trieste, Italy\\
$^{8}$Department of Astronomy, University of Maryland, College Park, MD 20742, USA\\
$^{9}$Harvard Smithsonian Center for Astrophysics, 60 Garden Street, Cambridge, MA 02138, USA\\
$^{10}$Dipartimento di Matematica e Fisica, Universit\`{a} degli Studi Roma Tre, via della Vasca Navale 84, I-00146 Roma, Italy
}

\date{Accepted XXX. Received YYY; in original form ZZZ}

\pubyear{2016}

\begin{document}
\label{firstpage}
\pagerange{\pageref{firstpage}--\pageref{lastpage}}

\maketitle

% Abstract of the paper
\begin{abstract}

We present an analysis of the joint {\it XMM-Newton} and {\it NuSTAR} observations of the radio-quiet quasar QSO\,B2202--209. Using an optical observation from the Hale Telescope at the Palomar Observatory, we revise the redshift of the source from the previously reported $z=1.77$ to $z=0.532$, and we estimate the mass of the central black hole, $\log (M_{\rm BH}/M_{\odot}) = 9.08 \pm 0.18$. The X-ray spectrum of this source can be well described by a power-law of photon index $\Gamma = 1.82 \pm 0.05$ with $E_{\rm cut} = 152_{-54}^{+103}\,{\rm keV}$, in the rest frame of the source. Assuming a Comptonisation model, we estimate the coronal temperature to be $kT_{\rm e}=42\pm 3 \,{\rm keV}$ and $kT_{\rm e}= 56 \pm 3\,{\rm keV}$ for a spherical and a slab geometry, respectively. The coronal properties are comparable to the ones derived for local AGN, despite a difference of around one order of magnitude in black hole mass and X-ray luminosity ($L_{2-10} = 1.93\times 10^{45}\,{\rm erg\,s^{-1}}$).
The quasar is X-ray loud, with an unusually flat observed optical-to-X-ray spectral slope $\alpha_{\rm OX} = 1.00 \pm 0.02$, and has an exceptionally strong optical [\ion{O}{iii}] line. Assuming that both the X-ray emission and the [\ion{O}{iii}] line are isotropic, these two extreme properties can be explained by a nearly edge-on disk, leading to a reduction in the observed UV continuum light.

\end{abstract}

% Select between one and six entries from the list of approved keywords.
% Don't make up new ones.

%%%%%%%%%%%%%%%%%%%%%%%%%%%%%%%%%%%%%%%%%%%%%%%%%%
%%%%%%%%%%%%%%%%% BODY OF PAPER %%%%%%%%%%%%%%%%%%

\section{Introduction}
\label{sec:intro}

It is generally thought that the primary hard X-ray continuum emission arising around accreting black holes (active galactic nuclei (AGN) and X-ray binaries), is due to Compton up-scattering of UV/soft X-ray disc photons off a hot ($\sim 10^9\,{\rm K}$), trans-relativistic medium, usually referred to as the X-ray corona \citep[e.g.][]{Shap76, Haa93, Pet01a, Pet01b}. The resulting spectrum, known as the `primary' emission, can be described by a power-law with a high-energy exponential cutoff that depends on the Compton $y$-parameter of the medium, which is a function of the electron temperature ($kT_{\rm e}$) and the optical depth ($\tau$) of the corona. These electrons are thought to be heated and confined by magnetic fields emerging from the ionized accretion disc \citep[e.g.][]{Gal79, Haa91, Mer01}.

In addition to being directly detected by an observer at infinity, the primary emission will irradiate and be `reflected' by the accretion disc \citep{Geo91}. X-rays incident on the disc will be subjected to Compton scattering and photoelectric absorption followed either by Auger de-excitation or by fluorescent line emission \citep[see e.g.][]{Light88, Geo91}. The resulting reflection spectrum is characterized by the iron K$\alpha$ emission line at $\sim 6.4-7.0\,{\rm keV}$ (depending on the ionization state of the disc) and a broad component peaked at around 20--30\,keV, known as `Compton hump'. It should be noted that the accretion disc is not the only reflector able to reprocess the primary emission in AGN. Many sources reveal the presence of narrow Fe K$\alpha$ emission lines \citep[e.g.][]{Bia09} that could be explained by reflection from distant, dense material such as the narrow line region or the putative molecular torus invoked in AGN unification models \citep{Ghisel94}.

Several lines of evidence suggest that the corona is compact, and located close to the black hole. Spectral-timing and reverberation studies, for example, are suggestive of a physically small corona that lies $3-10\,r_{\rm g}$ (where $r_{\rm g} = GM/c^2$ is the gravitational radius of a BH with mass $M$) above the central BH \citep[e.g.][]{Fab09,Emma14,Par14,Gal15}. It has been inferred from X-ray microlensing analyses of some bright lensed quasars that the hard X-rays are emitted from compact regions with half-light radii less than 6$\,r_{\rm g}$ \citep{Mor08,Char09,Reis13}. Moreover, eclipses of the X-ray source have also placed constraints on the size of the hard X-ray emitting regions, $r \lesssim 10\,r_{\rm g}$ \citep[e.g.][]{Ris07,Mai10, San13}. Analyses of the emissivity profile of the accretion disc \citep{Wil11,Wil12} suggest an extended corona a few gravitational radii from the accretion disc. More recently, \cite{Wil15a} and \cite{Wil15b} explained the long and short time-scale variabilities seen in Mrk\,335 in the context of an ``aborted jet launch'' as proposed earlier by \cite{Ghis04}. However, the coronal temperature and optical depth have remained poorly constrained due to the lack of high-quality X-ray measurements extending above 10 keV. Complex spectral components, including reflection from the accretion disc as well as from distant matter, contribute to the whole X-ray spectrum, making a precise determination of the cutoff energy challenging.

The unprecedented sensitivity of {\it NuSTAR} \citep{Har13} covering the $3-79\,{\rm keV}$ band, has provided for the first time the capability to measure spectral parameters in a precise and robust way including both the Compton reflection component and the high-energy cutoff ($E_{\rm cut} \sim 100-180\,{\rm keV}$) in some of the brightest local AGN, with luminosities on the order of $10^{42-44}\,{\rm erg\,s^{-1}}$ in the {\it NuSTAR} energy band \citep[e.g.][]{Mar14,Bren14,Bal14,Balo15,Matt15}. Combining the high sensitivity of {\it XMM-Newton} \citep{Jans01} at soft X-ray energies with that of {\it NuSTAR} at hard energies, we are able to investigate the coronal properties of more distant and more luminous quasars. This allows to understand better the coronal physics and the dependence of the corona on luminosity and redshift.

In this work, we study the X--ray spectrum provided by a coordinated {\it XMM-Newton} and {\it Nustar} observation of the radio-quiet quasar (RQQ) QSO\,B2202--209 (hereafter B2202, also known as PB\,5062). The redshift of the source was estimated by \cite{Reb87} to be $z=1.77$, implying a luminosity of $L_{2-10} \simeq 3\times 10^{46}\,{\rm erg\,s^{-1}}$, which would make this source erroneously, as we will show, the most luminous RQQ within its redshift \citep{Eitan13}. However, by analysing its X-ray spectrum, we found the need of an additional absorber at an intermediate redshift of 0.53. \cite{Reb87} determined the redshift of the source by identifying a broad emission line observed at $\sim 4290\,$\AA with a \ion{C}{iv}$\lambda 1549$ line. However, instead identifying the line with \ion{Mg}{ii}$\lambda 2800$ implies a redshift of 0.532, compatible with the redshift of the X-ray absorber. This discrepancy motivated us to obtain a higher quality optical spectrum using the Hale Telescope at Palomar Observatory, presented in the next section. The Palomar data shows that the source is actually situated at a redshift of 0.532. The main aim of this project is to measure the coronal properties of the source. The measurements of $E_{\rm cut}$ for local radio-quiet AGN are based on the deviation (on the order of 10-30\%) from a power-law at the highest energy part of the {\it NuSTAR} spectrum. Considering the redshift of B2202, particularly for the previously assumed $z=1.77$, $E_{\rm cut}$ is shifted to lower observed energies, helping compensate for the cosmological dimming.

This paper is structured as follows: Section\,\ref{sec:obsred} describes the observations and the data reduction; Section\,\ref{sec:SpecAnal} presents the spectral analysis; Finally, in Section\,\ref{sec:discussion} we provide a brief summary of our main results and discuss their implications. The following cosmological parameters are assumed: $\Omega_{\rm M} = 0.27$, $\Omega_{\Lambda} = 0.73$, and $H_0 = 70\,{\rm km\,s^{-1}\,Mpc^{-1}}$.

\section{Observations and data reduction}
\label{sec:obsred}

\subsection{Palomar Observations}
\label{subsec:Optical}

We obtained an optical spectrum of B2202 using the dual-beam
Double Spectrograph on the 200-inch Hale Telescope at Palomar
Observatory.  The 900\,s spectrum, obtained on UT 2016 May 28 in
photometric conditions, used the 1\farcs5 wide slit, the 5500\,\AA\
dichroic to split the light, the $600\,\ell\,{\rm mm}^{-1}$ grating on
the blue arm ($\lambda_{\rm blaze} = 4000$\,\AA; spectral resolving
power $R \equiv \lambda / \Delta \lambda \sim 1200$), and the 
$316\,\ell\,{\rm mm}^{-1}$ grating on the red arm ($\lambda_{\rm blaze} =
7500$\,\AA; $R \sim 1800$).  We processed the data using standard
techniques within IRAF, and calibrated the spectrum using an
archival sensitivity function obtained in February 2016 using the
same instrument configuration and observing conditions.

\begin{figure*}
\centering
\includegraphics[width = 0.8\textwidth]{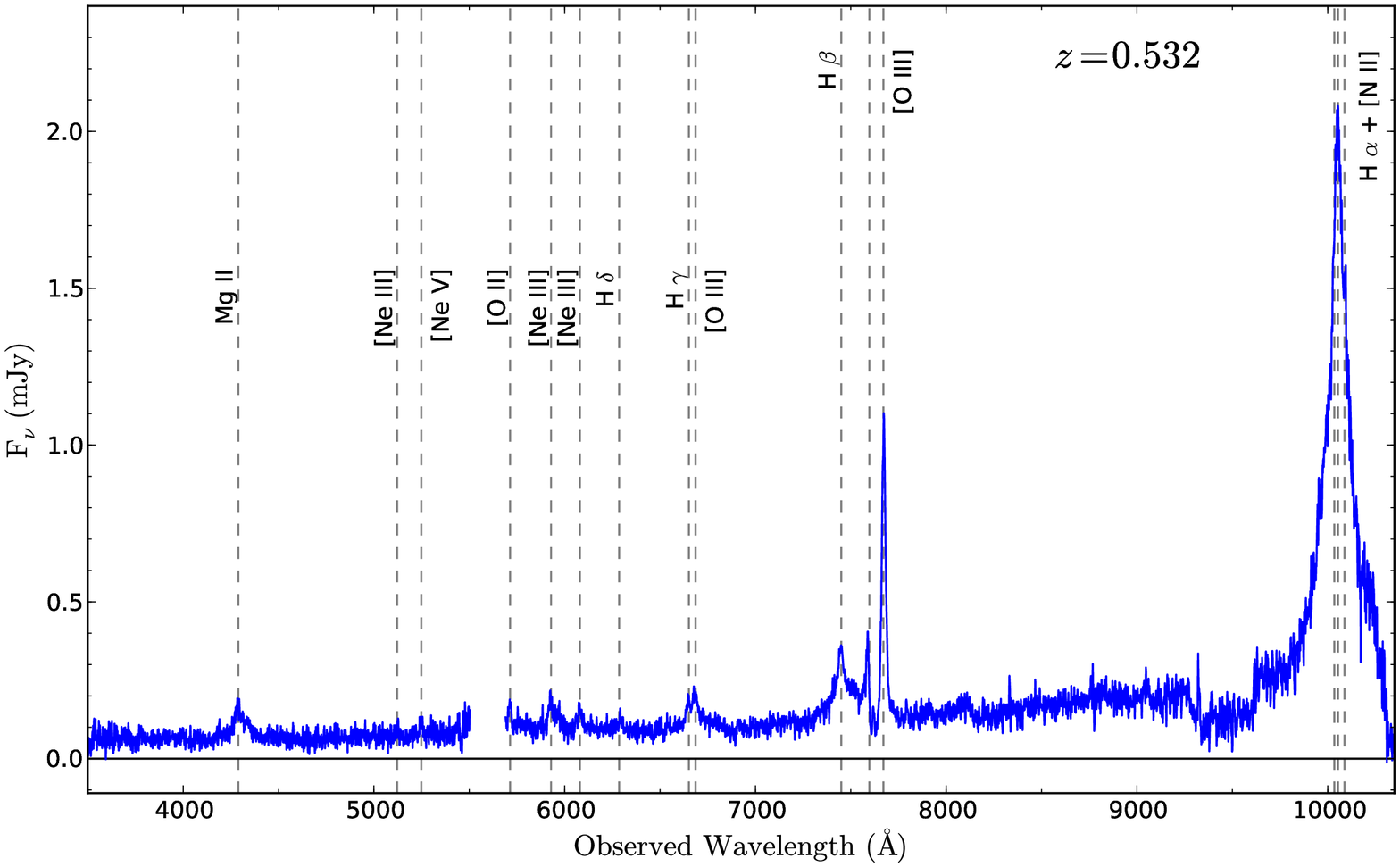}\\
\includegraphics[width = 0.9\textwidth]{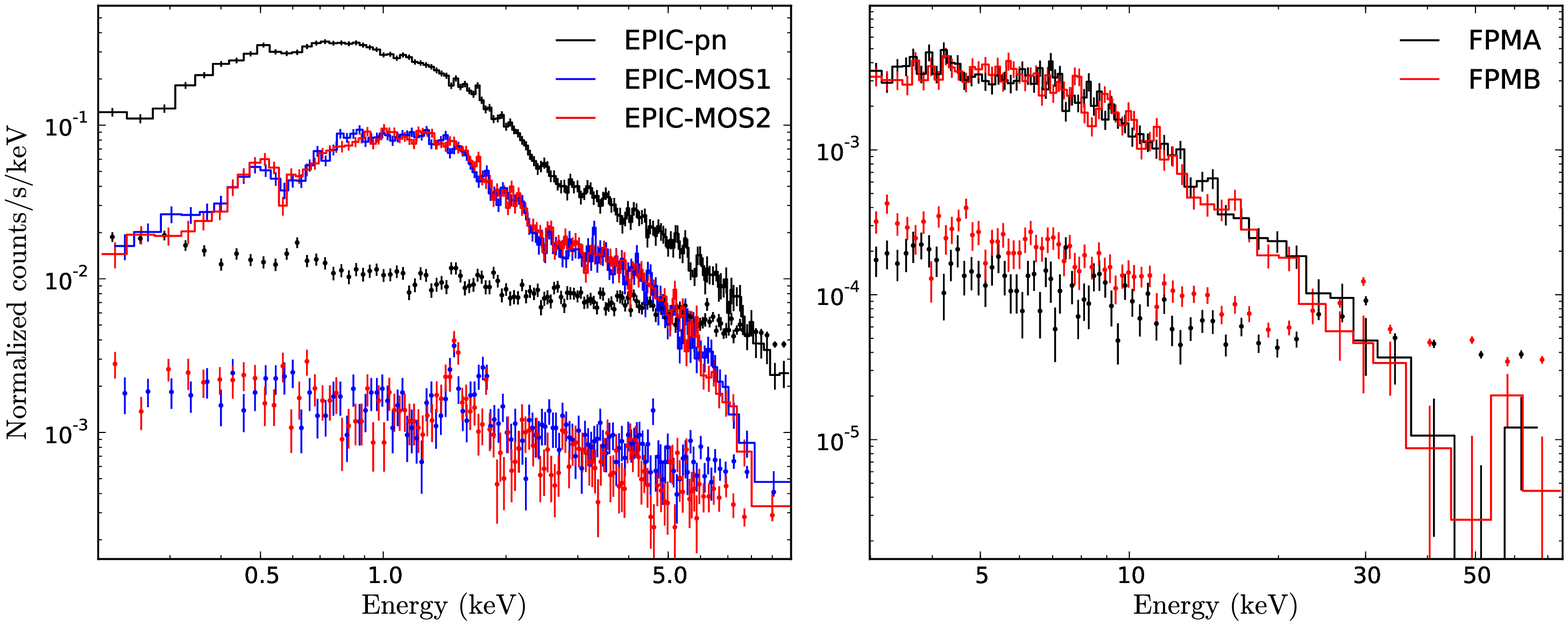}
\caption{Upper panel: optical spectrum obtained using the Palomar DBSP. Lower panels: X-ray background-subtracted source spectra (continuous lines) and background spectra (dots). Left panel: the spectra extracted from the EPIC-pn (black), EPIC MOS1 (blue) and EPIC MOS2 (red) instruments. Right panel: the spectra extracted from the FPMA (black) and FPMB (red) modules.}
\label{fig:Spec}
\end{figure*}

The processed spectrum, shown in the upper panel of Figure\,\ref{fig:Spec}, 
shows the source to be an AGN at $z = 0.532$ with many standard emission features identified,
including broad \ion{Mg}{ii}$\lambda 2800$, broad hydrogen Balmer
emission lines, and narrow [\ion{O}{iii}]$\lambda\lambda 4959, 5007$;
the redshift has been determined from the latter features. In particular, the 
new data clearly demonstrate that the line at $\sim 4290$\,\AA previously identified 
as \ion{C}{iv} by \cite{Reb87} is indeed \ion{Mg}{ii}.

%%%%%%%%%%%%%%%%%%%%%%%%%%%%%%%%%%%
%%%%%%%%%%%FIG-1

%%%%%%%%%%%%%%%%%%%%%%%%%%%%%%%%%%%%%%%

\subsection{X-ray observations}
\label{subsec:Xray}

B2202 was observed simultaneously by {\it XMM-Newton} and {\it NuSTAR}, on 2015 November 06-07 (Obs. IDs 0764370201 and 60101030002, respectively). The basic observation details are presented in Table\,\ref{table:log}. Here we summarise our data reduction procedures.

\subsubsection{XMM-Newton}
\label{subsec:XMMdata}
The {\it XMM-Newton} data were reduced using {\tt SAS v.15.0.0} and the latest calibration files.We followed the standard procedure for reducing the data of the EPIC-pn \citep{Stru01} and the two EPIC-MOS \citep{Tur01} CCD cameras, all operating in full frame mode with a thin filter for the EPIC-pn and a medium filter for the EPIC-MOS cameras. The EPIC-pn and EPIC-MOS data were processed using {\tt EPPROC} and {\tt EMPROC}, respectively. Source spectra and light curves were extracted from a circular region of radius $\sim 40\arcsec$. The corresponding background spectra and light curves were extracted from an off-source circular region located on the same CCD chip, with a radius approximately twice that of the source  to ensure a high signal-to-noise ratio. We filtered out periods with strong background flares estimated to be around 7.5\,ks. We corrected the light curves for the background count rate using {\tt EPICLCCORR}. The extracted light curves did not show any significant spectral variability. Response matrices were produced using the FTOOLs {\tt RMFGEN} and {\tt ARFGEN}. We re-binned the observed spectra, shown in the bottom left panel of Figure\,\ref{fig:Spec}, using the SAS task {\tt SPECGROUP} to have a minimum S/N of 5 in each energy bin. The MOS1 and MOS2 spectra are consistent, so we combined them using the SAS command {\tt COMBINE}.

%%%%%%%%%%%%%%%%%%%%%%%%%%%%
%%%%%%%%%%%%%% TABLE 2
\begin{table}
\centering
\caption{Net exposure times, and count rates estimated from the background-subtracted data in the 0.5--10 keV range for {\it XMM-Newton}, and the 4-30 keV range for {\it NuSTAR}.}
\begin{tabular}{ccc}
\hline \hline

Instrument	&	Net exposure 	&			Count Rate	\\[0.2cm] 
	&	 (ks)	&			 (count/s)\\ \hline 
EPIC-PN	&	60.5	&	$	0.501	\pm	0.003	$	\\[0.2cm]
MOS\,1	&	69.7	&	$	0.152	\pm	0.001	$	\\[0.2cm]
MOS\,2	&	69.4	&	$	0.156	\pm	0.001	$	\\[0.2cm]
FPMA	&	106.8	&	$	0.0233	\pm	0.0005	$	\\[0.2cm]
FPMB	&	106.5	&	$	0.0227	\pm	0.0005	$	\\[0.2cm] \hline\hline

\end{tabular}
\label{table:log}
\end{table}

%%%%%%%%%%%%%%%%%%%%%%%%%%%%%%%%%%%%%%%%%%%%%%%%%%%%%%%%%%%%%%%%%%%%%%%%%%%%%%%%%%%%%%%%%%%%

%%%%%%%%%%%%%%%%%%%%%%%%%%%%%%%%%%%%%%%%
%%%%%%%%%%%%%%%%%%%%%%%%%%%%%%%%%%%%%%%%

\subsubsection{NuSTAR}
\label{subsec:Nustardata}

We reduced the {\it NuSTAR} data following the standard pipeline in the {\it NuSTAR} Data Analysis Software (NuSTARDAS v1.4.1), and instrumental responses from {\it NuSTAR} CALDB v20151008. We cleaned the unfiltered event files with the standard depth correction, which significantly reduces the internal background at high energies, and  South Atlantic Anomaly passages were excluded from our analysis. We extracted the time-averaged source and background spectra from circular regions of radii $40\arcsec$ and $100\arcsec$, respectively, for both focal plane modules (FPMA and FPMB) using the HEASoft task {\tt NUPRODUCT}, and requiring a minimum of 50 counts per bin. The extracted spectra are presented in the bottom right panel of Figure\,\ref{fig:Spec}. The spectra extracted from both modules are consistent with each other. The background starts to dominate the source above $\sim 30$\,keV. For that reason, we decided to analyse the {\it NuSTAR} data in the observed $4-30$\,keV  energy range, which corresponds to the $6-46$\,keV energy range in the rest frame of the source. The data from FPMA and FPMB are analysed jointly in this work, but they are not combined together.

\section{Spectral Analysis}
\label{sec:SpecAnal}

Throughout this work, spectral fitting was done using {\tt XSPEC v12.9} \citep{Arn96}. Unless stated otherwise, uncertainties are listed at the 90\% confidence level ($\Delta \chi^2 = 2.71$). We included a variable constant, for each instrument, in order to account for the residual uncertainties in the flux calibration between the various detectors, fixing the constant for the EPIC-pn data to unity. We considered the EPIC-pn and the merged EPIC-MOS spectra in the 0.5--10\,keV range, and the FPMA/B spectra in the 4-30 keV band, as mentioned above.

%%%%%%%%%%%%%%%%%%%%%%%%%%%%%%%%%%%%%%%%%%%%%%%%%%%
%%%%%%%%%%%%%%%%%%%%%%%%Figure 2
\begin{figure}
\centering
\includegraphics[width = 0.45\textwidth]{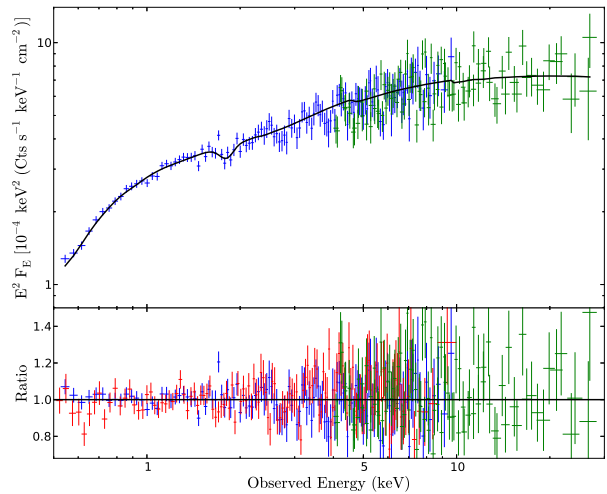}\\

\caption{Upper panel: observed spectra from EPIC-pn (blue) and {\it NuSTAR} (green) plotted together with the best-fit model (black line) composed of an absorbed cutoff power-law. Lower panel: data/model ratio for the EPIC-pn (blue), EPIC-MOS (red), and {\it NuSTAR} (green) data.}
\label{fig:modelratio}
\end{figure}

%%%%%%%%%%%%%%%%%%%%%%%%%%%%%%%%%%%%%%%%%%%%%%%%%%%
%%%%%%%%%%%%%%%%%%%%%%%%%%%%%%%%%%%%%%%%%%%%%%%%%%%

\begin{figure}
\centering
\includegraphics[width=0.45\textwidth]{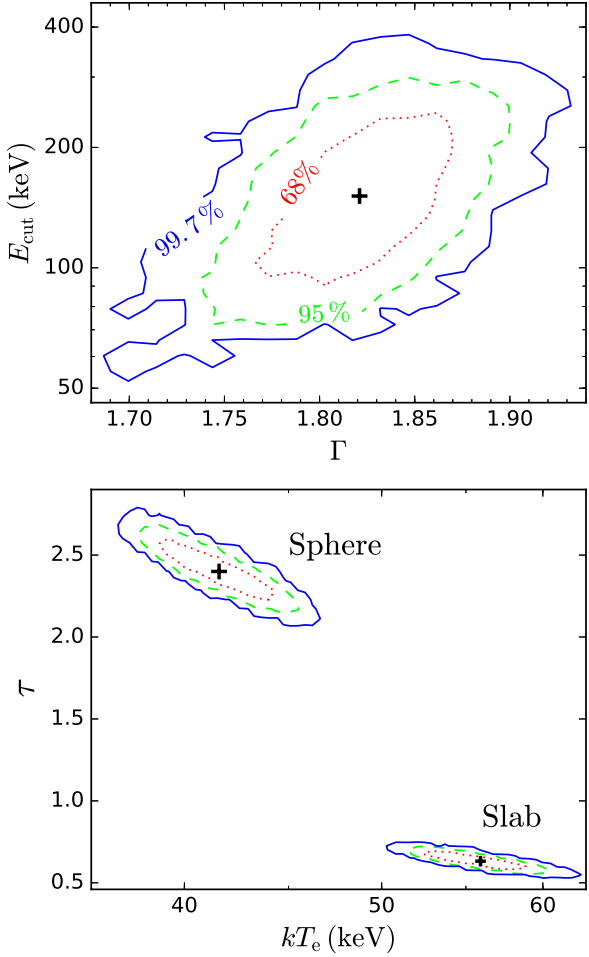}

\caption{Upper panel: $\Gamma-E_{\rm cut}$ contour plot (in the rest frame) for the cutoff powerlaw  model. Lower panel: $kT_{\rm e}-\tau$ contour plot (in the rest frame) for the Comptonisation model assuming a spherical geometry  and a slab geometry. We plot the 68\% (dotted red lines), 95\% (dashed green lines ) and 99.7\% (solid blue lines) confidence levels.}
\label{fig:cont}
\end{figure}

First, we fitted the spectra with a simple, absorbed power-law with an exponential high-energy cutoff ({\tt CUTOFFPL} model in {\tt XSPEC}). We fixed the Galactic absorption to the equivalent hydrogen column density in the line of sight towards B2202, $N_{\rm H} = 5.52\times 10^{20}\,{\rm cm^{-2}}$ \citep{Kal05}. Furthermore, we considered an intrinsic neutral absorber at the redshift of the source ({\tt ZWABS}). The model fit is unacceptable ($\chi^2/\nu =  1.32$), mostly due to the need of an extra ionised absorption component. For that reason, we add a partial covering absorption by partially ionized material ({\tt ZXIPCF}) at the redshift of the source. The fit improved ($\chi^2/\nu = 1.12$). However, a significant deficit could be detected in the residuals at energy $\sim 1.8$\,keV, most probably due to inaccuracies in modelling the Si absorption in the CCD detectors. We modelled this deficit using a Gaussian absorption profile with a free centroid energy, width, and normalisation (in the observed frame). The fit improves significantly by adding this line ($\Delta \chi^2/\Delta \nu = 25/3$, F-test null probability $=4.14\times 10^{-5}$), resulting in a $\chi^2/\nu = 1.05$. The final model and data/model ratio are presented in Figure\,\ref{fig:modelratio}. We performed a Monte Carlo Markov Chain (MCMC) analysis to estimate the errors on the parameters, using the Goodman-Weare algorithm \citep{Goo10} with a chain of 50,000 elements, and discarding the first 5000 elements as part of the ``burn-in" period. We found a high-energy cutoff $E_{\rm cut} = 99_{-35}^{+67}\,{\rm keV}$ in the observed frame that corresponds to an energy of $152_{-54}^{+103}\,{\rm keV}$ in the rest frame of the source, with a photon index $\Gamma = 1.82 \pm 0.05$ . In the upper panel of Figure\,\ref{fig:cont}, we present the contour plots derived from the MCMC analysis showing the constraints on the photon index $\Gamma$ and the high-energy cutoff $E_{\rm cut}$ in the rest frame of the source. Moreover, we identified a partially ionised absorber $\left( \log \left[ \xi({\rm erg\,s^{-1}\,cm^{-1}})\right] = 0.39_{-0.75}^{+0.80} \right)$ situated at the redshift of the source covering $\sim 30\%$ of the source and having a column density $N_{\rm H} = 2.32_{-0.88}^{+0.83}\times 10^{23}{\rm cm^{-2} }$. In addition, we identified a neutral absorber at the rest frame of the source with a column density $N_{\rm H} = 1.4_{-0.20}^{+0.22} \times 10^{21} {\rm cm^{-2} }$. The best-fit parameters are reported in Table\,\ref{table:Bestfit}.

%%%%%%%%%%%%%%%%%%%%%%%%%%%%%%%%%%%%%%%%%%%%%%%%%%
%%%%%%%%%%Table A2
\begin{table}
\centering
\caption{Best-fit parameters obtained assuming a cutoff power-law model, and a Comptonisation model for both spherical and slab geometries. }

\begin{threeparttable}
\begin{tabular}{cccc}
\hline \hline

Parameter	&				Cutoff						&				Compton						&				Compton						\\[0.2cm]   
	&										&				(spherical)						&				(slab)						\\[0.2cm]   \hline \\[-0.2cm]
\multicolumn{4}{c}{\tt zwabs} \\[0.2cm]																															
$N_{\rm H} \, (10^{21}\,{\rm cm^{-2}})$	&	$	1.40	_{-	0.20	}	^{+	0.22	}	$	&	$	1.40	_{-	0.12	}	^{+	0.13	}	$	&	$	1.38	_{-	0.13	}	^{+	0.15	}	$	\\[0.2cm] \hline \\[-0.2cm] 
\multicolumn{4}{c}{\tt zxipcf}	\\[0.2cm]																														
$N_{\rm H} \, (10^{23}\,{\rm cm^{-2}})$	&	$	2.32	_{-	0.88	}	^{+	0.83	}	$	&	$	2.32	_{-	0.55	}	^{+	0.68	}	$	&	$	2.43	_{-	0.45	}	^{+	0.77	}	$	\\[0.2cm]
$\log \left[ \xi({\rm erg\,s^{-1}\,cm^{-1}})\right]$	&	$	0.39	_{-	0.75	}	^{+	0.80	}	$	&	$	0.40	_{-	0.47	}	^{+	0.79	}	$	&	$	0.36	_{-	0.35	}	^{+	0.76	}	$	\\[0.2cm]
CF (\%)	&	$	30	_{-	5	}	^{+	6	}	$	&	$	30	_{-	3	}	^{+	4	}	$	&	$	29	_{-	3	}	^{+	4	}	$	\\[0.2cm] \hline \\[-0.2cm] 
\multicolumn{4}{c}{\tt cutoffpl}		\\[0.2cm]																													
$\Gamma$	&	$	1.82	\pm	0.05					$	&				--						&				--						\\[0.2cm]
$E_{\rm cut}^a\, ({\rm keV})$	&	$	99	_{-	35	}	^{+	67	}	$	&				--						&				--						\\[0.2cm]
${\rm Norm \,(\times 10^{-4})}$	&	$	5.26	_{-	0.41	}	^{+	0.54	}	$	&				--						&				--						\\[0.2cm] \hline \\[-0.2cm]  
\multicolumn{4}{c}{\tt compTT}	\\[0.2cm]																														
$kT_{\rm e}^b\,{\rm (keV)}$	&				--						&	$	42	\pm 3						$	&	$	56		\pm 3					$	\\[0.2cm]
$\tau$	&				--						&	$	2.40	_{-	0.16	}	^{+	0.18	}	$	&	$	0.63	_{-	0.05	}	^{+	0.06	}	$	\\[0.2cm]
${\rm Norm \,(\times 10^{-4}) }$	&				--						&	$	4.23	_{-	0.41	}	^{+	0.45	}	$	&	$	3.07	_{-	0.28	}	^{+	0.34	}	$	\\[0.2cm] \hline \\[-0.2cm]
\multicolumn{4}{c}{\tt gaussian (absorption)}	\\[0.2cm]																														
$E\, ({\rm keV})$	&	$	1.8	\pm	0.04					$	&	$	1.8	\pm	0.04					$	&	$	1.8	\pm	0.03					$	\\[0.2cm]
$\sigma\,({\rm eV})$	&	$	72	_{-	40	}	^{+	55	}	$	&	$	72	_{-	34	}	^{+	47	}	$	&	$	73	_{-	38	}	^{+	54	}	$	\\[0.2cm]
${\rm Norm \,(\times 10^{-6}) }$	&	$	4.75	_{-	2.34	}	^{+	1.77	}	$	&	$	4.73	_{-	1.87	}	^{+	1.47	}	$	&	$	4.79	_{-	2.17	}	^{+	1.58	}	$	\\[0.2cm] \hline \\[-0.2cm]
${\rm \chi^2/d.o.f.}$	&	$			363/344					$	&	$			363/344					$	&	$			363/344					$	\\[0.2cm]
	&	$			1.05					$	&	$			1.05					$	&	$			1.05					$	\\[0.2cm]\hline \hline

\end{tabular}

\begin{tablenotes}

\item {\bf Notes.}
\item[$^{\rm a}$] The cutoff energy is reported in the observed frame.
\item[$^{\rm b}$] The electron temperature is reported in the rest frame of the source.

\end{tablenotes}

\end{threeparttable}
\label{table:Bestfit}
\end{table}

%%%%%%%%%%%%%%%%%%%%%%%%%%%%%%%%%%%%%%%%%%%%%%%%%%

The measurement of $E_{\rm cut}$ could be affected by the presence of a reflection component in the spectrum. In order to test the stability of our measurement, we replaced the power-law model by the neutral reflection model {\tt PEXMON} \citep{Nan07}, considering an intermediate inclination of $60\degr$ and solar abundances. The fit improves by $\Delta \chi^2 = -6$ for one additional free parameter. We found a low reflection fraction ($R_{\rm frac} = 0.31_{-0.21}^{+0.24}$) and a low high-energy cutoff of $E_{\rm cut}=69_{-25}^{+47}\,{\rm keV}$ in the rest frame of the source, but consistent within the error bars with the value found assuming a simple cutoff power-law model. In addition, we tested the possibility of having relativistic reflection using the {\tt RELXILL} model \citep{Daus13,Gar14, Daus16}. The fit was statistically accepted ($\chi^2/\nu = 1.04$), and $E_{\rm cut}$ was consistent with the values determined above. However, neither the reflection fraction nor the ionisation parameter of the disc could be well constrained.

Finally, assuming that the power-law spectrum is obtained by Comptonisation of soft photons arising from the accretion disc by hot electrons in a corona, we substituted the cutoff power-law with the Comptonisation model {\tt COMPTT} \citep{Tit94}. This model allows us to determine the coronal temperature and optical depth assuming either a spherical or a slab geometry. We fixed the temperature of the seed photons, assumed to follow a Wien distribution law, to $kT=20\,{\rm eV}$. We obtained a fit statistically similar to the one obtained for the cutoff power-law model. We performed an MCMC analysis similar to the one mentioned previously, and found the coronal temperature and optical depth to be $kT_{\rm e}=42 \pm 3\,{\rm keV},\,\tau = 2.40_{-	0.16	}^{+	0.18	}$ (spherical geometry), $kT_{\rm e}=56 \pm 3\,{\rm keV},\, \tau= 0.63_{-0.05}^{+0.06}$ (slab geometry). The electron temperatures are in agreement with the estimation of $E_{\rm cut}$, within the error bars, obtained assuming a high-energy cutoff power-law model \citep[$E_{\rm cut} \simeq 2-3 kT_{\rm e}$, e.g.][]{Pet01b}. The optical depth is larger for the spherical geometry. This is expected because the optical depth estimated assuming a spherical geometry is the radial (effective) depth, while the slab optical depth is the vertical one that should be lower than the effective one. The $kT_{\rm e}-\tau$ contour plots for both geometries are presented in the lower panel of Figure\,\ref{fig:cont}. The best-fit parameters are shown in Table\,\ref{table:Bestfit}. The observed 2--10\,keV flux is $F_{2-10}=(1.43 \pm 0.07)\times 10^{-12}\,{\rm erg\,cm^{-2}\,s^{-1}}$, while the unabsorbed 2--10\,keV luminosity, in the rest frame, is $L_{2-10} = (1.93 \pm 0.07)\times 10^{45}\,{\rm erg\,s^{-1}}$.

B2202 was also observed by {\it XMM-Newton} in May 2001 (Obs. ID 12440301, hereafter obs.\,1), for $\sim 30\,{\rm ks}$. We extracted the EPIC-pn/MOS data in a similar way to that described in Section\,\ref{subsec:XMMdata}. We fit the EPIC spectra of this observation by an absorbed power-law model, similar to the best-fit model presented above, but without considering either the high-energy cutoff or the absorption feature at $\sim 1.8\,{\rm keV}$. The fit is good ($\chi^2/\nu = 0.97$) and the parameters of both neutral and ionised absorbers are consistent, within the error bars, with the ones found in 2015. However, we found a steeper power-law with a photon index $\Gamma = 1.97_{-0.06}^{+0.05}$. During obs.\,1, the source was $\sim 1.5$ times brighter than in 2015 (unabsorbed luminosity, $L_{2-10} = 2.92 \times 10^{45}\,{\rm erg\,s^{-1}}$) showing a behaviour often observed in AGNs, where the X-ray spectrum gets steeper when the source is brighter.

%%%%%%%%%%%%%%%%%%%%%%%%%%%%%%%%%%%%%%%%%%%%%%%%%%%55
%%%%%%%%%%%%%%%%%%%%%%%%Figure 3
%%%%%%%%%%%%%%%%%%%%%%%%%%%%%%%%%%%%%%%%%%%%%%%%%%%55
%%%%%%%%%%%%%%%%%%%%%%%%%%%%%%%%%%%%%%%%%%%%%%%%%%%55

\section{Discussion and conclusions}
\label{sec:discussion}

We present joint {\it XMM-Newton} and {\it NuSTAR} observations of the luminous quasar B2202. The 0.5--30\,keV spectrum of this source could be fitted with an absorbed (a neutral absorber and a partially ionised absorber at the redshift of the source) power-law ($\Gamma \sim 1.82$) with an exponential high-energy cutoff ($E_{\rm cut} \sim 153\,{\rm keV}$, at the rest frame of the source). No strong reflection features have been identified in the spectrum. 

In order to derive the black hole mass ($M_{\rm BH}$) of B2202, we fit the H$\beta$ line region with the IDL package MPFIT \citep[][]{Mark09}. Approximating the uncertainties to the flux using the standard deviation of the spectrum outside the emission lines, we fit the rest-frame 4200-5600\,\AA\ including a power-law continuum, broad and narrow Gaussian components for the H$\beta$ line, narrow [\ion{O}{iii}]$\lambda\lambda 4959, 5007$ doublet, narrow \ion{He}{ii}$\lambda 4686$ and the blended H$\gamma$ and [\ion{O}{iii}]$\lambda 4363$. The contribution from the \ion{Fe}{ii} complex is negligible for this target and thus was ignored. 
The full widths at half maximum (FWHM) and the equivalent widths (EW) of the H$\beta$ lines and the [\ion{O}{iii}]$\lambda\lambda 4959, 5007$ doublet are presented in Table\,\ref{table:elines}. 
Using the width of the broad H$\beta$ line and the monochromatic 5100\,\AA\ luminosity ($L_{5100}$), we get $\log (M_{\rm BH}/M_{\odot})=9.08 \pm 0.18$ from the \cite{Jun15} estimator. This implies an Eddington luminosity of $L_{\rm Edd} = (1.56 \pm 0.28)\times 10^{47}\,{\rm erg\, s^{-1}}$. Note that $L_{5100}$ is derived directly from the spectrum, that was checked to yield consistent i-band magnitude to that from the SDSS\footnote{\url{http://skyserver.sdss.org/dr10/en/home.aspx}} imaging, within a 10\% difference. We applied then the bolometric correction to $L_{5100}$ suggested by \cite{Marconi04}, $L_{\rm bol}/\nu_{\rm B}L_{\rm B} = 7.9 \pm 2.9$, and we estimate the bolometric luminosity of this source to be $L_{\rm bol}=(5.72 \pm 2.15)\times 10^{45}\,{\rm erg\,s^{-1} }$. This means that the source is accreting at $\sim 3.66\%$ of its Eddington limit. However, by applying the bolometric correction to $L_{2-10}$ \citep[equation 21 in][]{Marconi04}, we obtain a higher bolometric luminosity, $L_{\rm bol} = 1.79\times 10^{47}\,{\rm erg\,s^{-1}} \simeq 1.15\,L_{\rm Edd}$. 
Moreover, we estimated the monochromatic luminosity of the source at 2500\,\AA\ to be $L_{\rm \nu}(2500$\,\AA$ ) =  (1.06 \pm 0.02)\times 10^{45}\,{\rm erg\,s^{-1}}$, which implies an optical-to-X-ray spectral slope $\alpha_{\rm OX} = -0.384 \times \log[L_{\rm 2\,keV}/L_{2500}] =1.01 \pm 0.02$. This is a strong outlier in the $\alpha_{\rm OX}$-luminosity relation \citep[$\alpha_{\rm OX}\sim 1.6$ for typical quasars of the same luminosity,][]{Lusso16}.
However, the large EW of [\ion{O}{iii}]$\lambda 5007$ can be indicative of a high inclination angle $\theta$ between between the disc axis and the line of sight. In fact, we observe the projected EW along the line of sight, ${\rm EW_O =  EW^\ast} / \cos \theta $, where ${\rm EW_O}$ is the observed EW, and ${\rm EW^\ast}$ is the EW as measured in a face-on disc. By considering an average value of ${\rm EW^\ast}$, for all quasars, to be $\sim 11$\,\AA \citep{Ris11,Bis16}, we obtain $ \theta \simeq 85^\circ $. This means that the disc is observed nearly edge-on and can explain the high X-ray loudness of the source. In other terms, having such a high inclination of the disc with respect to the line of sight, the intrinsic UV luminosity should be higher than the observed one. Instead, the inclination will not affect the X-ray primary that is thought to be emitted isotropically. If we assume an intrinsic $L_{5100}$ to be larger than the observed value by a factor of $1/\cos \theta$, then this leads to an intrinsic $\alpha_{\rm OX,\,int} \simeq 1.43$ comparable to other sources. In addition, using the [\ion{O}{iii}]$\lambda 5007$ luminosity ($L_{[\ion{O}{iii}]} = (2.07 \pm 0.32)\times 10^{43}\,{\rm erg\,s^{-1}}$), we estimated the expeected X-ray luminosity, following the correlation found by \cite{Pan06}, to be $L_{\rm X} = (3.2 \pm 0.6)\times 10^{45}\,{\rm erg\,s^{-1}} $. The estimated value of $L_{\rm X}$ is in agreement within a factor of 1.65 with the value derived from the X-ray analysis in Sec.\,\ref{sec:SpecAnal}. Given the isotropicity of the X-ray primary emission, the high inclination of the disc and the agreement between the X-ray and the [\ion{O}{iii}] measurements, this is suggestive that, for this source, the measurement of $L_{\rm bol}$ using the X-ray luminosity is more reliable compared to the bolometric correction estimated using $L_{5100}$.

%%%%
%%%%%%%%%%%%%%%%%%%%%%%%%%%
%%%%%%%%%%%%%% TABLE 3
\begin{table}
\centering
\caption{Full widths at half maximum and equivalent widths of the broad and narrow H$\beta$ lines and the [\ion{O}{iii}]$\lambda\lambda 4959, 5007$ doublet seen in the Palomar spectra.}
\begin{tabular}{ccc}
\hline \hline

Line	&		FWHM (${\rm km\,s^{-1} }$)				&	EW (\AA)				\\[0.2cm] \\ \hline \\[-0.2cm]
H$\beta_{\rm broad}$	&	$	6940	\pm	1033	$	& $	128.4	\pm	5.9	$	\\[0.2cm]
H$\beta_{\rm narrow}$	&	$	 961	\pm	49	$	&	$ 18.4	\pm	8.5	$	\\[0.2cm]
[\ion{O}{iii}]$\lambda 4959$	&	$	1144	\pm	77	$	&	$49	\pm	3	$	\\[0.2cm]
[\ion{O}{iii}]$\lambda 5007$	&	$	995	\pm	50	$	&	$ 145.7	\pm	3.4	$	\\[0.2cm]\hline

\end{tabular}
\label{table:elines}
\end{table}
%%%%
%%%%%%%%%%%%%%%%%%%%%%%%%%%
%%%%%%%%%%%%%%

%%%%%%%%%%%%%%%%%%%%%%%%%%%%%%%%%%%%%%%%%%%%%%%%%%%55
%%%%%%%%%%%%%%%%%%%%%%%%Figure 4
\begin{figure*}
\centering
\includegraphics[width=0.9\textwidth]{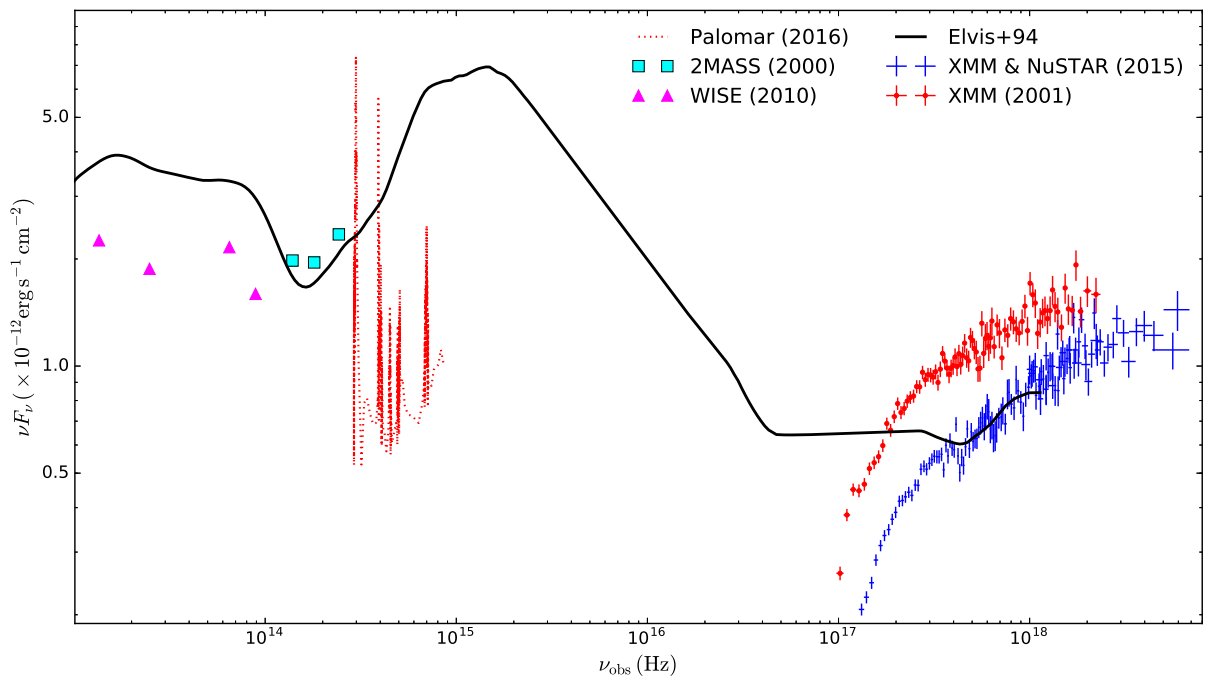}

\caption{The observed SED of the source, obtained using photometric data from {\it WISE} in 2010 (magenta triangles) and 2MASS in 2000 (cyan squares), in addition to the Palomar Observatory spectrum obtained in 2016 (red dotted line) and corrected for the Galactic extinction and for the Oxygen A-band absorption, the {\it XMM-Newton} spectrum obtained in 2001 (red dots) and the joint {\it XMM-Newton} and {\it NuSTAR} observations in 2015 (blue points). The typical radio-quiet quasar SEDs \citep{Elv94} is also plotted (black solid line) for comparison.}
\label{fig:SED}
\end{figure*}
%%%%%%%%%%%%%%%%%%%%%%%%%%%%%%%%%%%%%%%%%%%%%%%%%%%55

Interestingly, we found that the coronal properties are in agreement with the ones determined for local less luminous and less massive Seyfert galaxies \citep[e.g.][and references therein]{Fab15, Lub16}. \cite{Lub16} analysed the hard X-ray spectra of 28 Seyfert galaxies based on observations with {\it INTEGRAL} \citep{Win03}. The values of the electron temperature, photon index and reflection fraction that we found are in good agreement with the median values of their full sample ($\langle kT_{\rm e} \rangle = 48_{-14}^{+57}\,{\rm keV}$, $\langle \Gamma \rangle = 1.81_{-0.05}^{+0.18}$, and $ \langle R_{\rm frac} \rangle = 0.32_{-0.28}^{+0.33}$, respectively). 
\cite{Lub16} suggest that the small value of $R_{\rm frac}$ is indicative of a small solid angle under which the corona is seen from the disc. In other terms, this means a reduction in the flux of seed photons emitted by the disc that are cooling the corona. Moreover, they suggest the need of an additional process (such as synchrotron self-Compton) that is able to explain the efficient cooling of of a compact corona, close to the BH and separated geometrically from the accretion disc. 
%%%
Note that the low amount of reflection is in agreement with the X-ray Baldwin effect \citep[e.g.][]{Iwa93} whereby the equivalent width of the Fe\,K$\alpha$ line diminishes with X-ray luminosity. In fact, we modelled the Fe\,K$\alpha$ line by adding a Gaussian in emission, to the high-energy cutoff power-law model, and we estimated its EW to be $\leq 40\,{\rm eV}$. 
%%%
However, in our case, the high inclination of the disc can explain the small values of $R_{\rm frac}$ and  ${\rm EW(Fe\,K\alpha)}$, since most of the reflected light coming from the accretion disc will not be emitted in the direction of the observer. Instead, the weak reflection can be associated with sparse clouds, most likely to be in the broad line region, covering a small fraction of the solid angle. The similarities in both spectral shape and cut-off energy between B2202 and local AGNs are therefore indicative of a universal process determining the spectrum of the X-ray emission, with the strength of the coupling between disc and corona affecting only the overall normalization of such spectra.

We plot in Figure\,\ref{fig:SED} the observed multi-epoch spectral energy distribution (SED) of the source by adding, to the optical and X-ray spectra analysed in this paper, the {\it XMM-Newton} spectrum obtained in 2001 and the archival photometric data obtained from the Two Micron All Sky Survey\footnote{\url{http://irsa.ipac.caltech.edu/Missions/2mass.html}} \citep[2MASS, observed in 2000;][]{Skr06}, and the {\it Wide-field Infrared Survey Explorer}\footnote{\url{http://irsa.ipac.caltech.edu/Missions/wise.html}}({\it WISE}, observed in 2010; \citealt{Wri10}). We compare the observed SED to a typical SED of a RQQ \citep{Elv94} trying to match the X-ray spectra obtained in 2015. Fitting the SED properly, taking into account the complications and the long time-scale variability of the source, is beyond the scope of this paper. However, it is obvious that the optical spectrum is fainter and redder compared to a standard SED, but this is expected given the high inclination of the accretion disc, which leads the emission from the host galaxy, usually negligible, to show up. Moreover, interestingly, the IR emission appears to be low compared to the standard quasar SED. Adding to this the fact that the X-ray spectra are barely affected by absorption, and the small amount of reflection, this implies a particular system in which we observe the accretion disc nearly edge-on with no evidence of any Compton thick pc scale reprocessor, indicating a small covering fraction by the torus, if any.

This source was chosen on the basis of the previously reported redshift that mistakenly let it be considered one of the most luminous RQQs. However, our results on a less extreme quasar demonstrate that joint {\it NuSTAR} and {\it XMM-Newton} observations with a moderate exposure time are capable of making good measurements of coronal properties for quasars at cosmological redshifts. This significantly extends the previous work done primarily on relatively local and lower luminosity Seyferts \citep[e.g.][]{Fab15, Lub16}.

\section*{Acknowledgements}

This research made use of data from the {\it NuSTAR} mission, a project led by the California Institute of Technology, managed by the Jet Propulsion Laboratory, and funded by NASA, {\it XMM-Newton}, an ESA science mission with instruments and contributions directly funded by ESA Member States and NASA, and the Hale Telescope at Palomar Observatory. This research has made use of the {\it NuSTAR} Data Analysis Software (NuSTARDAS) jointly developed by the ASI Science Data Center (ASDC, Italy) and the California Institute of Technology (USA).  We like to acknowledge S.G. Djorgovski for providing the Palomar observations. We thank the anonymous referee for comments and suggestions, which significantly contributed to improving the quality of the manuscript. EB received funding from the European Unions Horizon 2020 research and innovation programme under the Marie Sklodowska-Curie grant agreement no. 655324.
\paragraph*{}
Facilities: {\it NuSTAR}, Palomar: DBSP, {\it XMM-Newton}.

%%%%%%%%%%%%%%%%%%%%%%%%%%%%%%%%%%%%%%%%%%%%%%%%%%

%%%%%%%%%%%%%%%%%%%% REFERENCES %%%%%%%%%%%%%%%%%%

% The best way to enter references is to use BibTeX:

\bibliographystyle{mnras}
\bibliography{ek-B2202ref} % if your bibtex file is called example.bib

%%%%%%%%%%%%%%%%%%%%%%%%%%%%%%%%%%%%%%%%%%%%%%%%%%

%%%%%%%%%%%%%%%%% APPENDICES %%%%%%%%%%%%%%%%%%%%%

%\appendix

% Don't change these lines
\bsp	% typesetting comment
\label{lastpage}
\end{document}